\documentclass[12pt]{iopart}

\usepackage{graphicx} 
\usepackage{epstopdf}
\usepackage[numbers]{natbib}
\usepackage{units}
\usepackage{url}
\usepackage[usenames,dvipsnames]{xcolor}
\usepackage{tikz}
\usepackage{enumerate}
\usepackage{pdflscape}
\usetikzlibrary{shapes,arrows}

\expandafter\let\csname equation*\endcsname\relax
\expandafter\let\csname endequation*\endcsname\relax
\usepackage{amsmath,amssymb}

\begin{document}

\title{Predicting Non-Square 2D Dice Probabilities} 

\author{G A T Pender and M Uhrin}
\address{Department of Physics and Astronomy, University College London, Gower Street, London WC1E 6BT, United Kingdom}
\ead{george.pender@cantab.net}

\begin{abstract}
The prediction of the final state probabilities of a general cuboid randomly thrown onto a surface is a problem that naturally arises in the minds of men and women familiar with regular cubic dice and the basic concepts of probability.  Indeed, it was considered by Newton in 1664 \citep{newton}.  In this paper we make progress on the 2D problem (which can be realised in 3D by considering a long cuboid, or alternatively a rectangular cross-sectioned dreidel). 

For the two-dimensional case we suggest that the ratio of the probabilities of landing on each of the two sides is given by $\frac{\sqrt{k^2+l^2}-k}{\sqrt{k^2+l^2}-l}\frac{\arctan{\frac{l}{k}}}{\arctan{\frac{k}{l}}}$ where $k$ and $l$ are the lengths of the two sides.  We test this theory both experimentally and computationally, and find good agreement between our theory, experimental and computational results.  

Our theory is known, from its derivation, to be an approximation for particularly bouncy or ``grippy" surfaces where the die rolls through many revolutions before settling.  On real surfaces we would expect (and we observe) that the true probability ratio for a 2D die is a somewhat closer to unity than predicted by our theory.

This problem may also have wider relevance in the testing of physics engines.
\end{abstract} 

\maketitle

\section{Introduction}
\label{Introduction}
\subsection{The Problem}
\begin{figure}[ht]
\begin{center}
\includegraphics[totalheight=0.3\textheight]{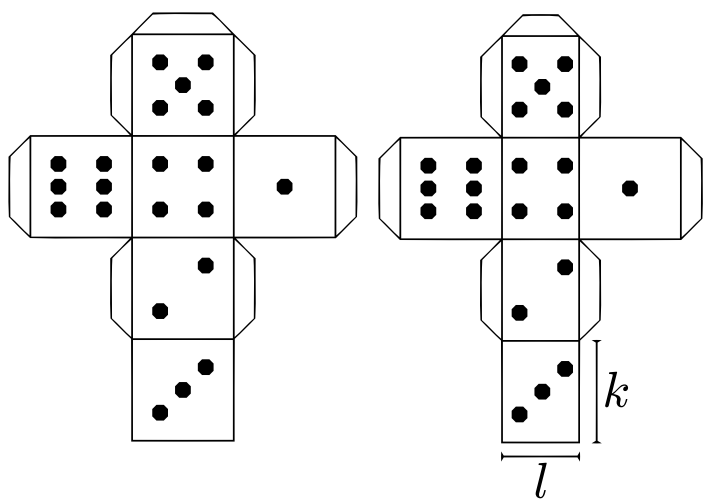}
\caption{Map for a fair die (left) and a die biased to give a higher variance (right).  Note that the average score for both dice would always be the same, owing to the rule that opposite sides of a die always sum to seven.}
\label{DiceMap}
\end{center}
\end{figure}

When a cube is thrown it bounces around and eventually has a one sixth chance of settling on any given side.  A deliberately biased die is usually made by modifying the weight distribution within a cube to alter the position of the centre of mass.  A die can also be biased if one of the dimensions is slightly longer or shorter than the others.  For example, the biased die (map shown on the right in figure \ref{DiceMap}) would have an enhanced probability of coming up with a six or a one, relative to a fair die (map on left).

\begin{figure}[ht]
\begin{center}
\includegraphics[totalheight=0.15\textheight]{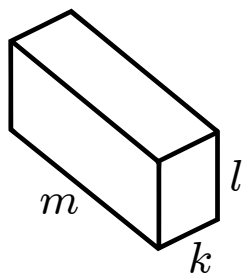}
\caption{General Question: What is the probability of the cuboid eventually settling with any face uppermost if it is given a random (large) initial angular momentum and a random (large) initial collision velocity with respect to the (reasonably rough and moderately elastic) ground?}
\label{GenCubDie}
\end{center}
\end{figure}

\begin{figure}[ht]
\begin{center}
\includegraphics[totalheight=0.2\textheight]{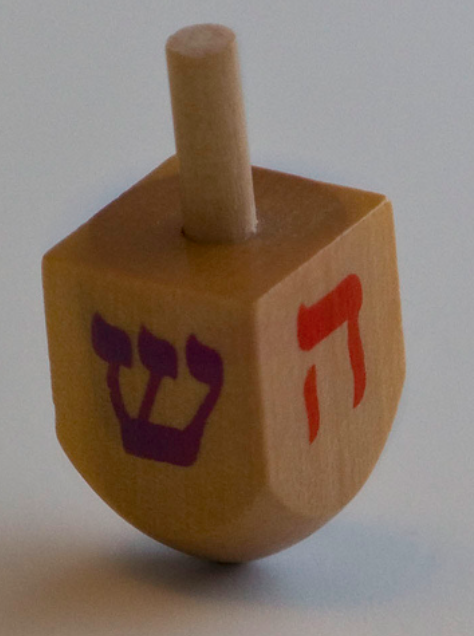}
\caption{Dreidel: A two-dimensional die which forms the basis of a traditional game played by Jewish children during the festival of Hanukkah.  Viewed from above, the dreidel has a square cross-section (giving a one quarter probability of landing on each side).  A biased dreidel could be made by using a rectangular (non-square) cross-section.  This is qualitatively understood and, for the same reason that traditional dice place the one and the six opposite each other, so too the best and worst outcomes for the dreidel (``gimmel" and ``shin") also tend to be placed opposite one another.}
\label{Dreidel}
\end{center}
\end{figure}

Exactly how the probability of getting a six might depend upon the ratio of $k$ to $l$ (or any other parameters) is a problem that we would like to solve.  In this paper we focus on the two-dimensional case.  This is equivalent to the three-dimensional case of a cuboidal ``die", where the length is very much greater than the breadth (and width).  One example of a two-dimensional die is that of the dreidel (a traditional jewish spinning toy, see figure \ref{Dreidel}) which is, effectively, a four-sided, 2D die.

Newton originally considered this problem as long ago as 1664.  More recently, \citet{Riemer} considered this problem for the three-dimensional case.  \citeauthor{Riemer} developed a semi-empirical method (based loosely on ideas from thermodynamics) with a free parameter (analogous to temperature) which is set empirically.  This parameter allows Riemer to adjust for the fact that other variables have an impact on the final state probabilities.  For example the properties of the surface may (and our experimental results suggest, do) affect the relative probabilities of the various possible outcomes. Nonetheless, the value of Riemer's free parameter has no theoretical justification even, for example, in the limiting case of a bouncy, high friction surface: $e \to 1 , \mu = 1$ (where $e$ and $\mu$ are the coefficients of restitution and friction).  This is clearly a drawback of Riemer's theory.

\citet{Mungan} have proposed another semi-empirical model, mathematically different from but, in broad form, similar to that of \citeauthor{Riemer}  Mungan's theoretical model is fitted to/tested on the same historical 1980s data set \cite{xxydata} as that of \citeauthor{Riemer} and both fit this data set to within binomial errors.

Our model (which only covers the behaviour of ``long" cuboids or other effectively two-dimensional die) contains certain approximations (for example we assume the die bounces a great many times before it comes to a stop) but, within those approximations, it does not require an empirically set parameter.

\subsection{Pedagogical aspects}

This paper should be easily comprehensible to an undergraduate student of
physics or mathematics.  The paper has relevance to the teaching of physics, in particular our
theoretical derivation could be used as a simple example of Markov Chain
analysis. The derivation of our main theoretical result could
well be used on a Markov Chains examples sheet.  However, a
prior knowledge of Markov Chains is not required to understand our argument.  The paper may also have a minor use to help in the teaching of binomial
error analysis and the use of computational models and the benefits (and
drawbacks) of their use, relative to physical experiments.

Finally, the replication of either our experimental or our computational
results, together with an explanation of the main theoretical derivation,
could well be the basis of a very creditable, low-tech and inexpensive,
class coursework project or introduction to the experimental method for A
level/IB (or equivalent) students.

\section{2D Theory}
\label{Theory}

\begin{figure}
\begin{center}
\includegraphics[totalheight=0.15\textheight]{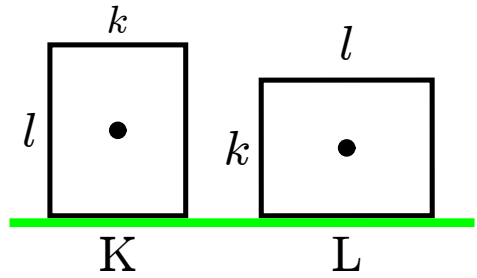}
\caption{There are two possible final states labelled ``K" and ``L".  The energy of these states are $E(K) = \frac{1}{2} mgl$ and $E(L) = \frac{1}{2} mgk$ respectively.}
\label{DiceStates}
\end{center}
\end{figure}

\begin{figure}
\begin{center}
\includegraphics[totalheight=0.2\textheight]{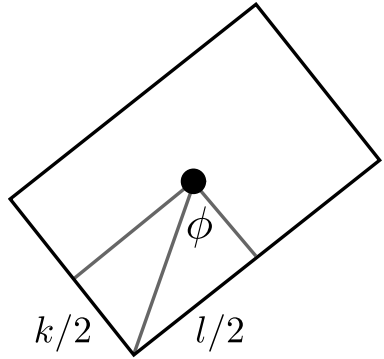}
\caption{If the die were dropped inelastically then the chance of landing in any particular orientation, $P_I\left(K\right)$ and $P_I\left(L\right)$, would simply be proportional to the angle subtended by that side of the die.}
\label{inelaslandchance}
\end{center}
\end{figure}

We consider the two state, two-dimensional, model shown in figure \ref{DiceStates}.  First suppose that the die is dropped (in a random orientation, with zero angular momentum and from a low height) and falls inelastically onto the floor.  Under these circumstances the probability of landing on any particular side would be proportional to the angle subtended by that side. For example, a die dropped from a low height in the orientation shown in figure \ref{inelaslandchance} would settle in state L (i.e. it would land on the long side).  Since all initial angular orientations must be equally likely, the die, if released in a random orientation and from a low height (still with zero initial angular momentum), would exhibit the following probabilities of landing in states K and L respectively:

\begin{equation}
P_I(K)=\frac{\arctan{\frac{k}{l}}}{\frac{\pi}{2}}\text{, }P_I(L)=\frac{\arctan{\frac{l}{k}}}{\frac{\pi}{2}}\label{InElasK}
\end{equation}

These probabilities are equivalent to those predicted by a two-dimensional version of what \citeauthor{Riemer} call the ``Simpson model" \citet{simpson}, named after Thomas Simpson who, in 1740, proposed that the probability of an $a\times b\times c$  cuboid landing on the $a \times b$ surface would be:

\begin{equation*}
P_{I}(\textrm{c-vertical})_{3D} = \frac{1}{\pi} \arctan \left( \frac{ab}{c\sqrt{a^2+b^2+c^2}} \right)
\end{equation*}

However, whichever way the die is dropped there is a probability of getting the other state due to the somewhat elastic nature of the collisions.  Consequently, the Simpson model tends to overestimate the probability of the higher energy outcomes.

To model bouncing from one state to the other (figure \ref{inelaslandchance}), we first have to recognise that to make the transition the block must have a total energy of at least

\begin{equation}
E_{Trans}=\frac{1}{2}mg\sqrt{l^2+k^2}\text{.}
\label{Etrans}
\end{equation}

\begin{figure}
\begin{center}
\includegraphics[totalheight=0.25\textheight]{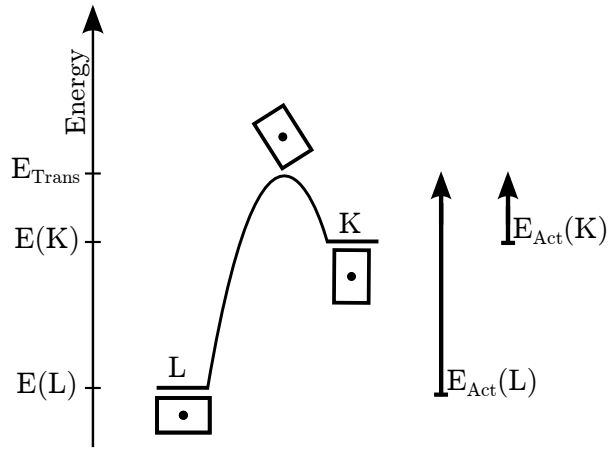}
\caption{Energy Level Diagram Showing two states and transition state.}
\label{EnergyLevDiag}
\end{center}
\end{figure}

To continue the analysis of this problem we consider a five state Markov Chain (figure \ref{MarkovChainCompleate}), the nodes of which are as follows:

\tikzstyle{block} = [rectangle,thick,draw,fill=blue!20,
    text width=6.2em, text centered, rounded corners, minimum
height=2.5em, node distance=4.2cm]
\tikzstyle{start_block} = [rectangle, draw, fill=red!20,
    text width=5em, text centered, rounded corners, minimum
height=2em, node distance=1cm]
\tikzstyle{line} = [draw, -latex', sloped]
\tikzstyle{plain_line} = [draw, -latex']

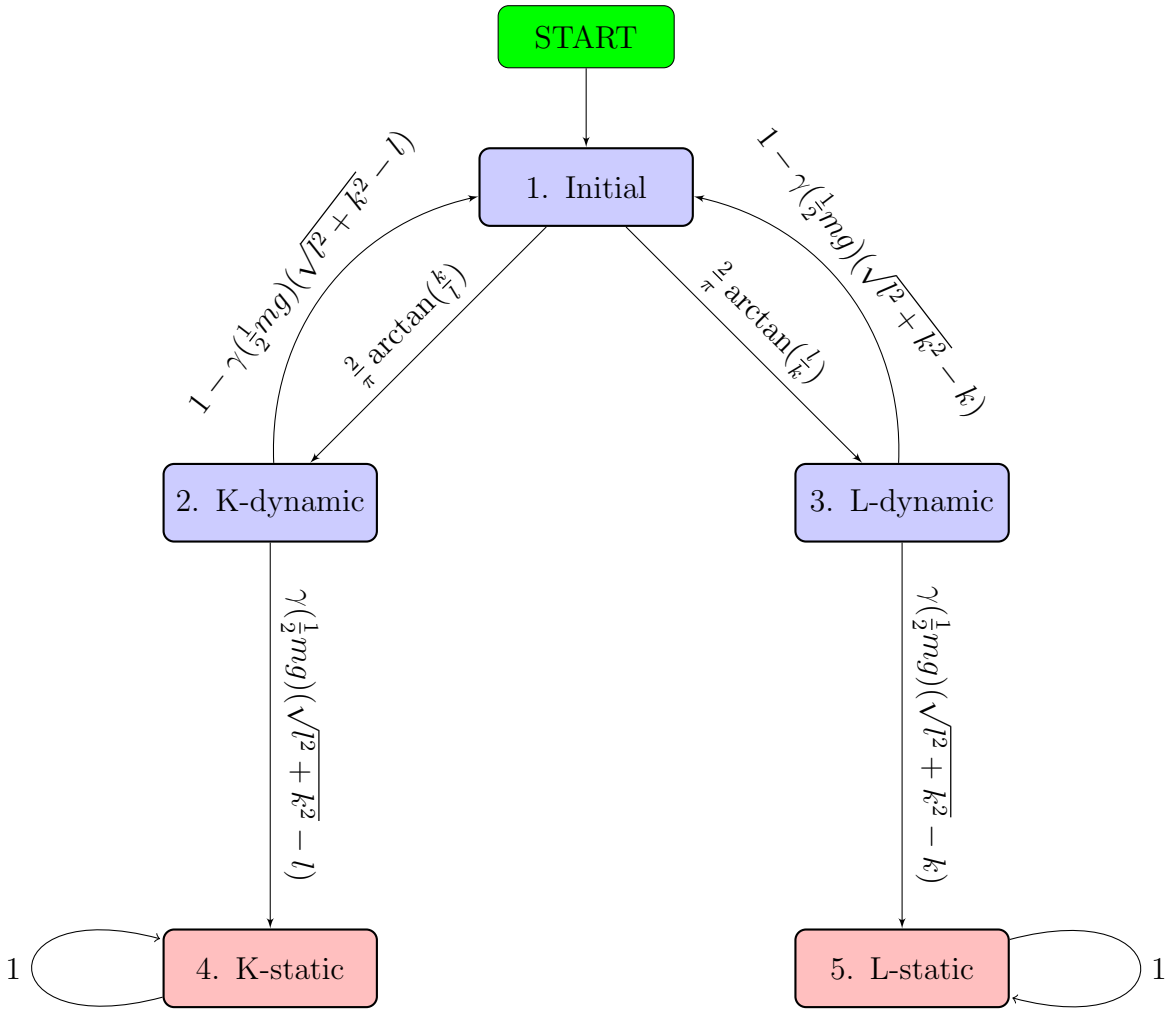
\begin{figure}
  \centering
\begin{tikzpicture}[auto]
    \node [start_block, fill=green] (start) {START};
    \node [block, below of=start, node distance=2cm] (drop) {1. Initial};
    \node [block, below of=drop, left of=drop] (k_dynamic) {2. K-dynamic};
    \node [block, below of=drop, right of=drop] (l_dynamic) {3. L-dynamic};
    \node [block, below of=k_dynamic, node distance=15em, fill=pink] (k_static) {4. K-static};
    \node [block, below of=l_dynamic, node distance=15em, fill=pink] (l_static) {5. L-static};
    \path [line] (start) -- (drop);
    \path [line] (drop) -- (k_dynamic) node[midway, above]
{$\frac{2}{\pi}\arctan(\frac{k}{l})$};
    \path [line] (k_dynamic) to[bend left=40] node[midway, above]
{$1-\gamma(\frac{1}{2}mg)(\sqrt{l^2+k^2}-l)$} (drop);
    \path [line] (k_dynamic) -- node[anchor=south]
{$\gamma(\frac{1}{2}mg)(\sqrt{l^2+k^2}-l)$} (k_static);
    \path [plain_line] (k_static) edge[loop left] node {1} ();
    \path [line] (drop) -- node[midway, above]
{$\frac{2}{\pi}\arctan(\frac{l}{k})$} (l_dynamic);
    \path [line] (l_dynamic) to[bend right=40] node[midway, above]
{$1-\gamma(\frac{1}{2}mg)(\sqrt{l^2+k^2}-k)$} (drop);
    \path [line] (l_dynamic) -- node[anchor=south]
{$\gamma(\frac{1}{2}mg)(\sqrt{l^2+k^2}-k)$}(l_static);
    \path [plain_line] (l_static) edge[loop right] node {1} ();
\end{tikzpicture}
  \caption{Final calculation of probability ratios for dice rolling
on a reasonably elastic and reasonably grippy surface with initial
angular momentum, initial hight and initial velocity picked from a
reasonably broad random distribution. $\gamma \to 0$ represents the
case where the dice bounce or roll a large number of times before
settling.}
\label{MarkovChainCompleate}
\end{figure}

\begin{enumerate}[1]
\item Top of figure \ref{MarkovChainCompleate}: A high-energy state (the initial state) where the die experiences a random chance of going into one or other dynamic state (states 2 and 3).  

\item Middle Left of figure \ref{MarkovChainCompleate}: A state (K-dynamic) where the die's centre of mass is over the side of length $k$ but it has enough energy to overcome the transition state (see figure \ref{EnergyLevDiag}) 

\item Middle Right of figure \ref{MarkovChainCompleate}: An analogous state (L-dynamic) where the centre of mass is over the side of length $l$ but, again, with a total energy of at least $E_{Trans}$ 

\item Bottom Left of figure \ref{MarkovChainCompleate}: A stable state (K-static) where the die is bound to end up in K because the die is nearly in final state K and has energy less than $E_{Trans}$.  

\item Bottom Right of figure \ref{MarkovChainCompleate}: An analogous stable state, ``L static".
\end{enumerate}

The probability arcs exiting the initial state are just the inelastic probabilities from equation \ref{InElasK}.  The probabilities exiting the two dynamic states are initially unknown but we know the exiting arcs from each dynamic state must sum to unity.  The chance of returning to the initial state will be higher for more elastic collisions and will tend to unity for highly elastic collisions.  This chance will be higher from K-dynamic than from L-dynamic as the lower bound on the energy of K-dynamic is higher than the lower bound on the energy of L-dynamic.  Once the system reaches either of the static states it remains in that state (indicated by the circular arcs found at nodes 4 and 5).

The key to solving the 2D problem lies in the assigning of the relative weights (in the limit of nearly elastic collisions) of the arcs leading from nodes 2 and 3.  We will assume that the die starts with a large amount of kinetic energy (both rotational and translational) and that this energy is gradually lost until the total energy is less than $E_{Trans} = \frac{1}{2} mg \sqrt{l^2+k^2}$, at which point the die settles.  The probability of getting from \textbf{L-Dynamic} to \textbf{L-Static} is therefore proportional to the ``activation energy", $E_{Act}(L)$, for this transition.

\begin{equation}
E_{Act}(L)=E_{Trans}-E(L)=\frac{1}{2}mg\left(\sqrt{l^2+k^2}-k \right)
\end{equation}

So the probabilities of the two ``downward" transitions (in figure \ref{MarkovChainCompleate}) are given by:

\begin{equation}
P=(\textrm{L-Dynamic} \to \textrm{L-Static}) = \gamma \left( \frac{1}{2} m g \right) \left( \sqrt{l^2+k^2}-k \right)
\end{equation}

\begin{equation}
P=(\textrm{K-Dynamic} \to \textrm{K-Static}) = \gamma \left( \frac{1}{2} m g \right) \left( \sqrt{l^2+k^2}-l \right)
\end{equation}

where $\gamma$ is a single (small) unknown. Numbering the five possible states as in figure \ref{MarkovChainCompleate}, from 1 to 5, the transpose of the transition matrix $\underline{\underline{P}}$ is therefore:

\begin{equation*}
 \underline{\underline{P}}^T= \left(
 \begin{matrix} 0&1-\gamma(\frac{1}{2}mg)(\sqrt{k^2+l^2}-l)&1-\gamma(\frac{1}{2}mg)(\sqrt{k^2+l^2}-k) & 0 & 0 \\
 \frac{2}{\pi}\arctan(k/l) & 0 & 0 & 0 & 0 \\
 \frac{2}{\pi}\arctan(l/k)&0&0&0&0 \\ 0&\gamma(\frac{1}{2}mg)(\sqrt{k^2+l^2}-l) & 0 & 1 & 0 \\
 0 & 0 & \gamma(\frac{1}{2}mg)(\sqrt{k^2+l^2}-k) & 0 & 1
 \end{matrix} \right)
\label{markovmatrix}
\end{equation*}

Denoting the elements of $\underline{\underline{P}}$ by $p_{ij}$ then, after $n$-steps, with an initial state vector $(1,0,0,0,0)^T$, we arrive at state $(\underline{\underline{P}}^T)^n (1,0,0,0,0)^T$.  In principle, it would be possible to calculate the final state probabilities $P(K)$ and $P(L)$ by recursively applying $\underline{\underline{P}}^T$ to the initial state vector for like so,

\begin{eqnarray*}
{\left( \underline{\underline{P}}^T \right)}^\infty \left( \begin{matrix} 1\\0\\0\\0\\0 \end{matrix} \right) = \left( \begin{matrix} 0\\0\\0\\P(K)\\P(L) \end{matrix} \right)\text{,}
\label{recur}
\end{eqnarray*}

however, it will be instructive to consider the state achieved after just two iterations. 

Applying $\underline{\underline{P}}^T$ to the initial state vector twice, gives:

\begin{eqnarray*}
 {\left(\underline{\underline{P}}^T\right)}^2 \left( \begin{matrix} 1\\0\\0\\0\\0 \end{matrix} \right) =\left( \begin{matrix} p_{12}p_{21}+p_{13}p_{31} \\0\\0\\p_{12}p_{24}\\p_{13}p_{35}\end{matrix} \right)
\end{eqnarray*}

Using the exact expression for $\underline{\underline{P}}^T$, and since $\arctan k/l + \arctan l/k = \pi/2$, we get for $\gamma \ll 1$:

\begin{eqnarray*}
 {\left(\underline{\underline{P}}_{\gamma \to 0}^T\right)}^2 \left( \begin{matrix} 1\\0\\0\\0\\0 \end{matrix} \right) = \left( \begin{matrix} \approx 1 \\0\\0\\p^{(2)}(1 \to 4)\\p^{(2)}(1 \to 5)\end{matrix} \right) =\left( \begin{matrix} \approx 1 \\0\\0\\\frac{\gamma m g}{\pi}(\sqrt{k^2+l^2}-l)\arctan(k/l)\\\frac{\gamma m g}{\pi}(\sqrt{k^2+l^2}-k)\arctan(l/k)\end{matrix} \right)
\end{eqnarray*}

Since all probability weight returned to the initial state will eventually be distributed to the final states in the same $p^{(2)}(1 \to 4):p^{(2)}(1 \to 5)$ ratio, we find that:

\begin{equation}
\frac{P(L)}{P(K)}=\frac{p^{(2)}(1 \to 5)}{p^{(2)}(1 \to 4)}= \frac{\sqrt{k^2+l^2}-k}{\sqrt{k^2+l^2}-l}.\frac{\arctan{\frac{l}{k}}}{\arctan{\frac{k}{l}}}
\label{MasterEquM1}
\end{equation}

Alternatively, it may be more intuitive to think in terms of the angle between the side of the block and the block diagonal (equivalent to the angle, $\phi$, in figure \ref{inelaslandchance}).

\begin{equation*}
\arctan \frac{l}{k} = \phi\text{, }\arctan \frac{k}{l} = \frac{\pi}{2}-\phi
\end{equation*}

In which case equation \ref{MasterEquM1} becomes:

\begin{equation}
\frac{P(L)}{P(K)}= \frac{1-\cos\phi}{1-\cos \left( \frac{\pi}{2}-\phi \right) }.\frac{\phi}{\frac{\pi}{2}-\phi}= \frac{\sin^2\frac{\phi}{2}}{\sin^2 \left( \frac{\pi}{4}-\frac{\phi}{2} \right) }.\frac{\phi}{\frac{\pi}{2}-\phi}
\end{equation}

When analysing experimental results it is easier to measure side length than to measure angles.  So we will plot experimental results in terms of side length ratios, $R=\frac{l}{k}$:

\begin{equation}
\frac{P(L)}{P(K)} = \frac{\sqrt{1+R^2}-1}{\sqrt{1+R^2}-R}\frac{\arctan{R}}{\arctan{\frac{1}{R}}}
\label{MasterEqu}
\end{equation}

Clearly $P(L)+P(K)=1$ and so the raw probabilities can be extracted from the probability ratio as follows:

\begin{equation}
P(K)= \frac{1}{1+\frac{P(L)}{P(K)}}
\label{rattoprob}
\end{equation}

\begin{figure}[ht]
\begin{center}
\includegraphics[totalheight=0.2\textheight]{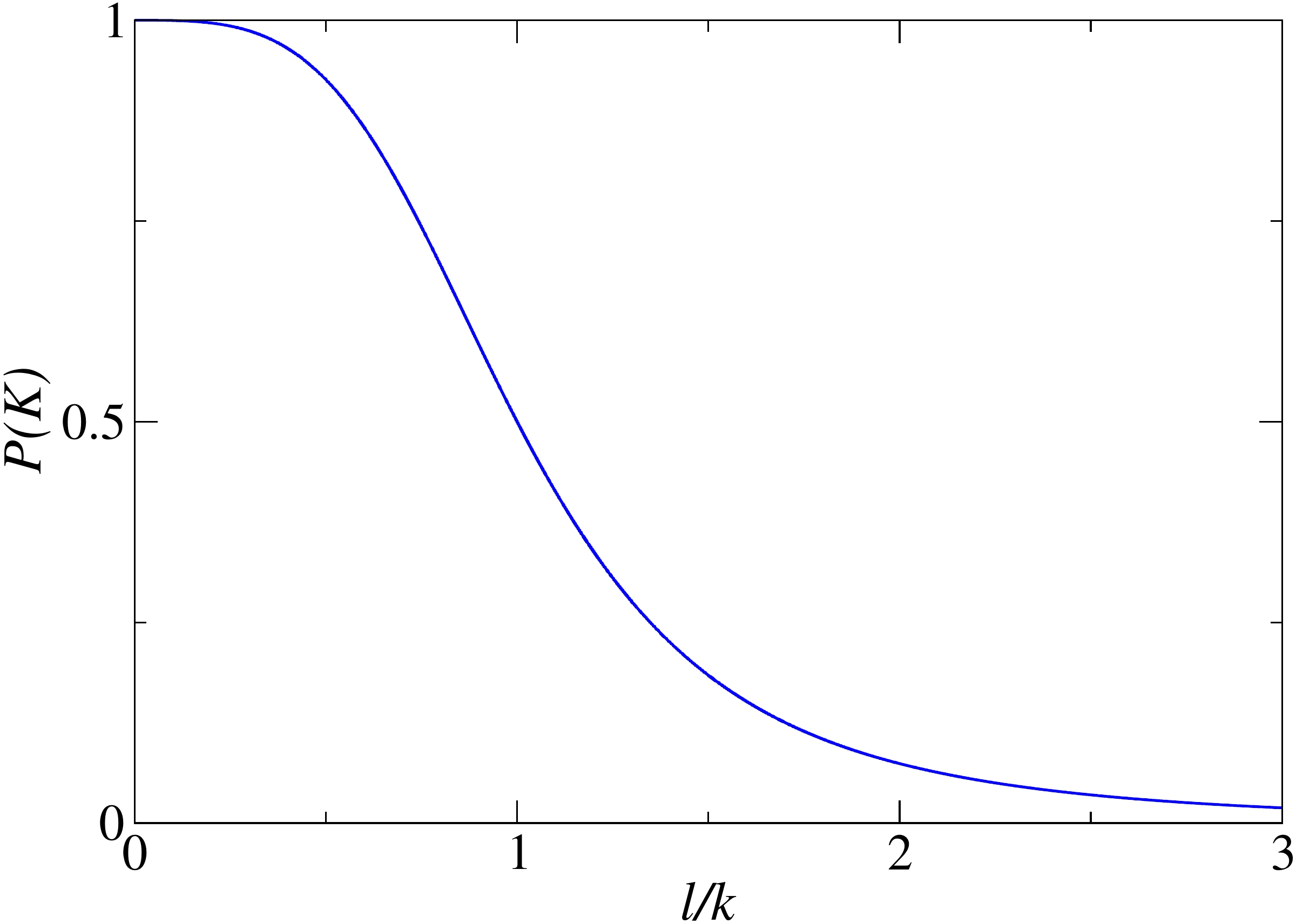}
\caption{{\bf2D Probability:} Here we show the predicted face probabilities for a two-dimensional block, although we will normally present our results as a probability ratio.}
\label{2Dprob}
\end{center}
\end{figure}

This probability is plotted in figure \ref{2Dprob}, however, generally we will present our results in terms of probability ratios (as per figure \ref{MasterGraph2}).  The use of ratios has three main advantages over the use of raw probabilities: it makes the expression mathematically tidier, it will later allow us to ignore the very small number of times the block lands in the third ``end-on" orientation, similarly it will allow us to more easily ignore the very small proportion of computational runs where the code yields an error of some kind.

The above derivation has tended to assume $ 1 \leq R < \infty$ we can cover exactly the same parameter space by relabelling the sides to have  $ 0 < R \leq 1$.  This makes it possible to plot the whole parameter space on an axis of finite length.



For more realistic surfaces (e.g. not highly bouncy) we would expect the frequency ratio to be closer to unity than predicted by equation \ref{MasterEqu}, while never being as close to unity as predicted by the ``no bounce" model (equation \ref{InElasK}). It is useful to to consider the role of $\gamma$ in our model.  
We initially though of it as being inversely linked to the ``bounciness" of the surface because decreasing it increases the probability that a die will continue to roll as opposed to settling on the current side.  Clearly there is an upper bound on $\gamma$ beyond which the back probabilities from the dynamic states become negative but up to this limit our final state probabilities are independent of its value. Our physical intuition therefore tells us that $\gamma$ is simply a dummy variable, not linked to any physical quantity and that, for surfaces of very low restitution, we should see probability predictions that are somewhat between our model and the ``no bounce" (or Simpson) model.

\section{Experiment - 2D}
\label{Experiment}

Wishing to test the derived relationship (equation \ref{MasterEqu}), two sets of long blocks were made. First, pine blocks were rolled on either carpet or a thin layer of towelling.  The second set were 3D printed from polylactide plastic using a MakerBot Replicator 2 with a quoted accuracy of 11  $\mu$m$^{\footnotemark}$\footnotetext{{\url{http://store.makerbot.com/replicator2.html}.  The 3D models used can be downloaded from https://github.com/muhrin/DicePhys/tree/master/3dmodels}}.  These were rolled, with initial angular momentum predominantly along the length, on a Medium Density Fibreboard (MDF) surface, as well as, separately, on tough carpet.

As this study is primarily concerned with two-dimensional blocks it is important to establish that the long side (length $m$) is sufficiently long that increasing it further does not affect the outcome of a roll.  Verification of this can be found in \ref{LengthApen} for several blocks.

\begin{figure}[ht]
\begin{center}
\includegraphics[totalheight=0.45\textheight]{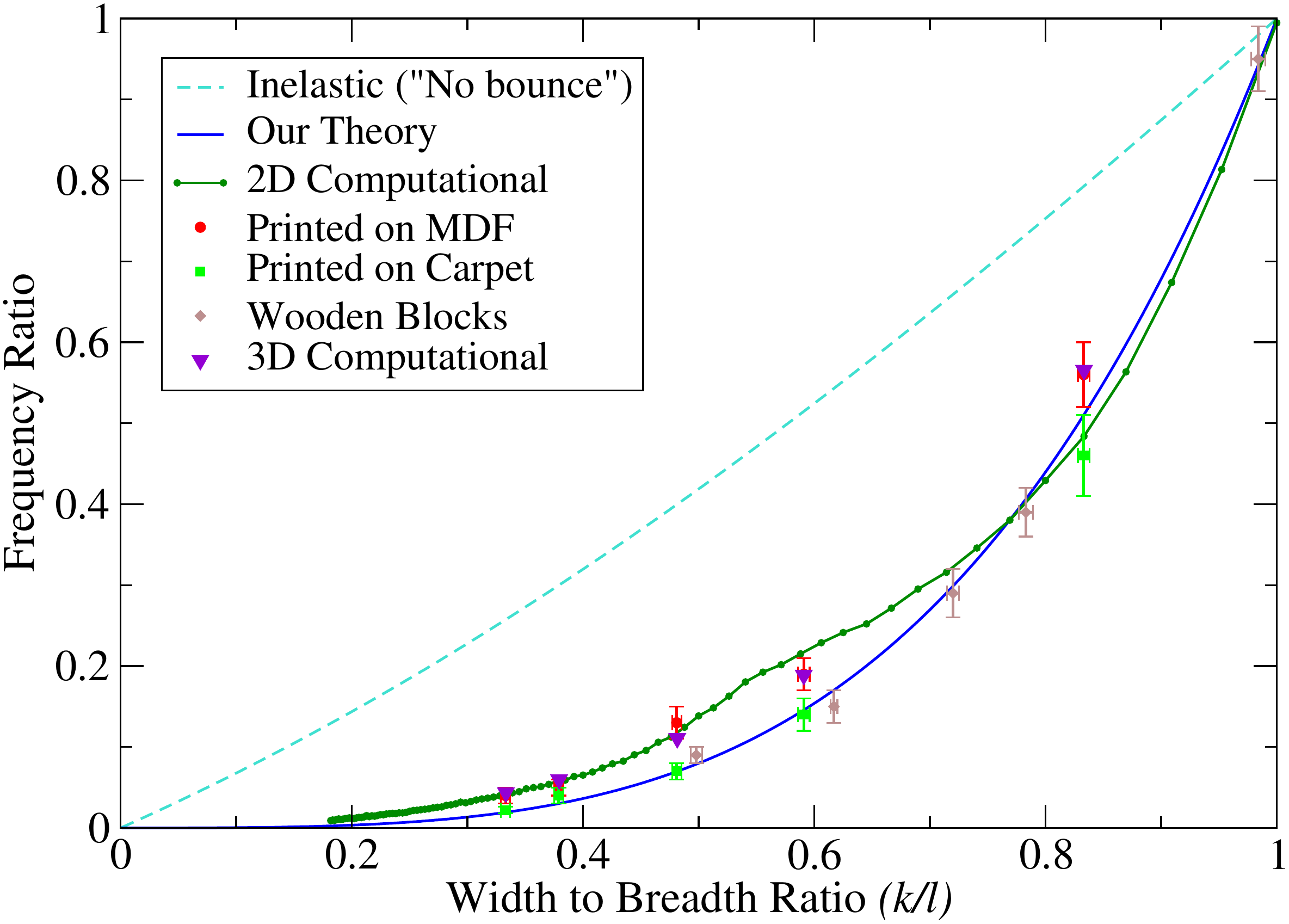}
\caption{Graph comparing theory from equation \ref{MasterEqu} (solid dark blue line) with experimental and computational results.  From our theory we would expect the ratio of final state frequencies (for long cuboids) to be very slightly above the theory line (formula given in equation \ref{MasterEqu}) with convergence for blocks which can be expected to roll many times before settling.  The dashed (lighter blue) line gives a ``No Bounce" prediction, the two-dimensional version of what \citeauthor{Riemer} \cite{Riemer} call the ``Simpson model".  This data represents between ten to twenty thousand individual experimental rolls.  Finally, the set of connected green circles and the purple triangles show results from our two- and three-dimensional computational simulations.}
\label{MasterGraph2}
\end{center}
\end{figure}

The full set of results are plotted in results figure \ref{MasterGraph2}.  Our theory makes good predictions for both wooden and printed blocks rolled on carpet, with a spread of data consistent with ordinary binomial errors. It is noticeable that the surface on which the block rolls has a systematic effect on the probability ratios. As expected, probability ratios measured for printed blocks on MDF are invariably closer to one than our theory would predict.  The same blocks rolled on carpet (a surface on which they roll, often through many revolutions, rather than ``clattering") give a probability ratio closer to our theoretical predictions.

Table \ref{sidewaystable} shows the full tabulation of all experimental results including details on the total number of rolls.  The probability ratio between the two outcomes, $L$ and $K$ (with measured frequency ratios $N_L$ and $N_K$, respectively), is estimated as:

\begin{equation*}
  \frac{N_L}{N_K} \pm \sqrt{\frac{N_L N_K \left( N_L + N_K \right)}{N_K^4}}
\end{equation*}

i.e. the probability is estimated, in the usual way, as:

\begin{equation*}
  \frac{N_L}{N_L+N_K} \pm \sqrt{\frac{N_L N_K}{\left( N_L+N_K \right)^3}}
\end{equation*}

\begin{landscape}
\begin{table}
\caption{Full Experimental Data. (Colours as per figure \ref{MasterGraph2})}
\label{sidewaystable}
\centering
\scriptsize
\begin{tabular}{|c c c|c c|c c|c c c c|c|}
\hline\hline
Length & Breadth & Width & Width & Predicted & Block & Surface & Number & Freq. & Freq. & Freq. & Freq. \\
(mm) & (mm) & (mm) & / Breadth & Probability & & & of & Length & Breadth & Width & Ratio $< 1$ \\ 
 &  & & $R$ & Ratio ($< 1$) & &&Throws & Vertical & Vertical & Vertical & binomial error \\ 
 [0.9ex] 
\hline
\color{RawSienna}$782 \pm 1$&\color{RawSienna}$25.0 \pm 0.1$&\color{RawSienna}$18.0 \pm 0.1$&\color{RawSienna}$0.720 \pm 0.005$&\color{RawSienna}$0.299 \pm 0.008$&\color{RawSienna}Pine&\color{RawSienna}Carpet&\color{RawSienna}800&&\color{RawSienna}178&\color{RawSienna}622&\color{RawSienna}$0.29 \pm 0.03$\\
$98 \pm 1$&$''$&$''$&$''$&$0.299 \pm 0.008$&$''$&$''$&200&&46&154&$0.30 \pm  0.06$\\
\color{RawSienna}$119.3 \pm 0.1$&\color{RawSienna}$61.7 \pm 0.4$&\color{RawSienna}$48.3 \pm 0.2$&\color{RawSienna}$0.783 \pm 0.006$&\color{RawSienna}$0.406 \pm 0.012$&\color{RawSienna}$''$&\color{RawSienna}Thin Towel&\color{RawSienna}800&\color{RawSienna}8&\color{RawSienna}222&\color{RawSienna}570&\color{RawSienna}$0.39 \pm 0.03$\\
\color{RawSienna}$360 \pm 1$&\color{RawSienna}$43.4 \pm 0.2$&\color{RawSienna}$21.6 \pm 0.2$&\color{RawSienna}$0.498 \pm 0.005$&\color{RawSienna}$0.079 \pm 0.003$&\color{RawSienna}$''$&\color{RawSienna}$''$&\color{RawSienna}1000&&\color{RawSienna}80&\color{RawSienna}920&\color{RawSienna}$0.09 \pm 0.01$\\
\color{RawSienna}$99.3 \pm 0.1$&\color{RawSienna}$43.9 \pm 0.2$&\color{RawSienna}$43.2 \pm 0.2$&\color{RawSienna}$0.984 \pm 0.006$&\color{RawSienna}$0.942 \pm 0.023$&\color{RawSienna}$''$&\color{RawSienna}$''$&\color{RawSienna}2000&\color{RawSienna}24&\color{RawSienna}962&\color{RawSienna}1014&\color{RawSienna}$0.95 \pm 0.04$\\
\color{RawSienna}$348 \pm 1$&\color{RawSienna}$70.3 \pm 0.2$&\color{RawSienna}$43.4 \pm 0.2$&\color{RawSienna}$0.617 \pm 0.003$&\color{RawSienna}$0.171 \pm 0.003$&\color{RawSienna}$''$&\color{RawSienna}Towel+Rug&\color{RawSienna}1000&&\color{RawSienna}130&\color{RawSienna}870&\color{RawSienna}$0.15 \pm 0.02$\\
\hline
\color{red}$140.0 \pm 0.1$&\color{red}$24.0 \pm 0.1$&\color{red}$20.0 \pm 0.1$&\color{red}$0.833 \pm 0.005$&\color{red}$0.511 \pm 0.013$&\color{red}Printed&\color{red}MDF&\color{red}1000&&\color{red}361&\color{red}639&\color{red}$0.54 \pm 0.04$\\
$100.0 \pm 0.1$&$''$&$''$&$''$&$0.511 \pm 0.013$&$''$&$''$&1000&&386&614&$0.63 \pm 0.04$\\
$80.0 \pm 0.1$&$''$&$''$&$''$&$0.511 \pm 0.013$&$''$&$''$&1000&4&342&654&$0.52 \pm 0.04$\\
\color{red}$140.0 \pm 0.1$&\color{red}$22.0 \pm 0.1$&\color{red}$13.0 \pm 0.1$&\color{red}$0.591 \pm 0.005$&\color{red}$0.146 \pm 0.005$&\color{red}$''$&\color{red}$''$&\color{red}1000&&\color{red}160&\color{red}840&\color{red}$0.19 \pm 0.02$\\
$80.0 \pm 0.1$&$''$&$''$&$''$&$0.146 \pm 0.005$&$''$&$''$&500&&84&416&$0.20 \pm 0.03$\\
\color{red}$140.0 \pm 0.1$&\color{red}$27.0 \pm 0.1$&\color{red}$13.0 \pm 0.1$&\color{red}$0.481 \pm 0.004$&\color{red}$0.070 \pm 0.002$&\color{red}$''$&\color{red}$''$&\color{red}500&&\color{red}58&\color{red}442&$\color{red}0.13 \pm 0.02$\\
\color{red}$''$&\color{red}$29.0 \pm 0.1$&\color{red}$11.0 \pm 0.1$&\color{red}$0.379 \pm 0.004$&\color{red}$0.030 \pm 0.001$&\color{red}$''$&\color{red}$''$&\color{red}500&&\color{red}26&\color{red}474&\color{red}$0.05 \pm 0.01$\\
\color{red}$''$&\color{red}$30.0 \pm 0.1$&\color{red}$10.0 \pm 0.1$&\color{red}$0.333 \pm 0.004$&\color{red}$0.019 \pm 0.001$&\color{red}$''$&\color{red}$''$&\color{red}500&&\color{red}19&\color{red}481&\color{red}$0.04 \pm 0.01$\\
\hline
\color{Green}$140.0 \pm 0.1$&\color{Green}$24.0 \pm 0.1$&\color{Green}$20.0 \pm 0.1$&\color{Green}$0.833 \pm 0.005$&\color{Green}$0.511 \pm 0.013$&\color{Green}Printed&\color{Green}Tough Carpet&\color{Green}500&&\color{Green}157&\color{Green}343&$\color{Green}0.46 \pm 0.05$\\
\color{Green}$''$&\color{Green}$22.0 \pm 0.1$&\color{Green}$13.0 \pm 0.1$&\color{Green}$0.591 \pm 0.005$&\color{Green}$0.146 \pm 0.005$&\color{Green}$''$&\color{Green}$''$&\color{Green}500&&\color{Green}62&\color{Green}438&\color{Green}$0.14 \pm 0.02$\\
\color{Green}$''$&\color{Green}$27.0 \pm 0.1$&\color{Green}$13.0 \pm 0.1$&\color{Green}$0.481 \pm 0.004$&\color{Green}$0.070 \pm 0.002$&\color{Green}$''$&\color{Green}$''$&\color{Green}500&&\color{Green}31&\color{Green}469&\color{Green}$0.07 \pm 0.01$\\
\color{Green}$''$&\color{Green}$29.0 \pm 0.1$&\color{Green}$11.0 \pm 0.1$&\color{Green}$0.379 \pm 0.004$&\color{Green}$0.030 \pm 0.001$&\color{Green}$''$&\color{Green}$''$&\color{Green}1000&&\color{Green}39&\color{Green}961&\color{Green}$0.04 \pm 0.01$\\
\color{Green}$''$&\color{Green}$30.0 \pm 0.1$&\color{Green}$10.0 \pm 0.1$&\color{Green}$0.333 \pm 0.004$&\color{Green}$0.019 \pm 0.001$&\color{Green}$''$&\color{Green}$''$&\color{Green}2000&&\color{Green}43&\color{Green}1957&\color{Green}$0.022 \pm 0.004$\\
\hline
\end{tabular}
\end{table}
\end{landscape}

\section{Computation}
\label{Computation}

Using a rigid body simulation, two sets of computational experiments were carried out: One constrained to two-dimensions and the other a direct analogue to our experiments, fully unconstrained in all three-dimensions, modelling the behaviour of the long cuboids thrown on MDF.
Our simulation code, DicePhys, uses the Bullet physics engine$^{\footnotemark}$\footnotetext{\url{http://bulletphysics.org/wordpress/}} to provide the rigid body dynamics and integration of equations of motion.  The full code is available online$^{\footnotemark}$\footnotetext{\url{https://github.com/muhrin/DicePhys}} and further details including convergence testing can be found in \ref{CompApen}.

The simulation world was configured to match the experimental conditions of the printed blocks on MDF as closely as possible.  Table \ref{tab:2d_ics} shows the set of initial conditions used throughout.  Where a range is shown a uniformly distributed random number spanning the interval was used.  The experimental procedure used to determine the frictional coefficient, $\mu$, is outlined in \ref{mu}.  The coefficient of restitution, $e$, was estimated to be 0.5.  For each set of dimensions a minimum of 100,000 virtual rolls were performed to achieve very small random errors in outcome probabilities.

\begin{table}[ht]
 \centering
  \begin{tabular}{|l|l|}
  \hline
  \hline
  Parameter & Initial value \\
  \hline
  Linear velocity (m/s) & $-0.2 \to 0.2$ \\
  Angular velocity (rads/s) & $31 \to 160$ \\
  Drop height (m) & $0.2 \to 0.4$ \\
  $\mu$ & 0.29 \\
  $e$ & 0.5 \\
  \hline
  \end{tabular}
  \caption{Initial conditions used in all simulations.}
  \label{tab:2d_ics}
\end{table}

The results of our computational simulations can be seen in figure \ref{MasterGraph2}.  Error bars for these curves are smaller than the size of the symbols and are therefore not shown.  All three-dimensional simulations (purple triangles) are in good agreement with the equivalent experimental results (red circles).  Agreement between experiment and the two-dimensional simulations (green circles connected by lines) is less good with the simulation line lying outside of two of the five experimental error bars.  This may be an artefact of the way the physics engine constrains the system to a plane when simulating in two dimensions as evidenced by the discrepancy between the two and three dimensional simulation results.

\section{The slightly off-square (moderately biased) dreidel}

Consider the case of a deliberately biased dreidel which has one side marginally longer than the other such that $\frac{l}{k} = R =1+\delta$ where $\delta \ll 1$.  With $\delta = 0.01$ using equation \ref{rattoprob}, $P(K) = 1/2.03737 \simeq 0.4908$.  Hence, the probability differs from that of a fair dreidel by $0.4908/0.5 \simeq 0.9817$, or close to 2\%.  To put it another way, each of the slightly longer sides would have a 25.46\% chance of being landed on as opposed to 24.54\% for either of the shorter sides.

Expanding equation \ref{2Dprob} around $R = 1$ gives a probability ratio of $\frac{P\left(L\right)}{P\left(K\right)}=1+\left( 1 + \sqrt{2} + \frac{4}{\pi} \right)\delta + O(\delta^2)$.  Combining this with equation \ref{rattoprob}, we see that an ordinary four sided dreidel, biased in this way, will now show side probabilities of $\frac{1}{4} \pm \frac{1+\sqrt{2}+\frac{4}{\pi}}{8} \delta$.

\section{Conclusions}

We have developed and tested (both experimentally and computationally) a
model for the landing probabilities of two-dimensional dice (or long 3D
cuboid, or biased dreidel).  We have found that, as suggested theoretically, the ratio or the
probabilities of landing in each of the two potential orientations (when
the plank is thrown onto a reasonably bouncy surface) is given by
$\frac{\sqrt{k^2+l^2}-k}{\sqrt{k^2+l^2}-l}\frac{\arctan{\frac{l}{k}}}{\arctan{\frac{k}{l}}}$
where $k$ and $l$ are the lengths of the two sides.

We hope that this result might be used in order to assist in the teaching
or Markov Chain analysis for young undergraduate students.

Unfortunately we have not managed to adapt this theoretical approach to the three-dimensional problem, this is an area for further thought.  It may be that our theory can be extended to three dimensions.  Alternatively it may be that, in three dimensions, the final state probabilities are dependent, in some complicated way, on the way in which the dice are thrown (for example, the outcome probabilities may be dependent on the typical magnitude and direction of the initial spin imparted to the die).  Although other work has been done which predicts three-dimensional, cuboidal dice probabilities (e.g. \citet{Riemer}, \citet{Mungan}), these models contain one or more free parameters which must be set experimentally (even in the limit of, for example, $e \to 1$, $\mu = 1$).

\section*{Acknowledgements}

The authors would like to thank the anonymous referee whose detailed comments helped to significantly improve the manuscript.  The authors would also like to thank the University College London Institute of Making for allowing us to use their 3D printers.

\bibliography{2Ddice}

\newpage
\appendix

\section{Determining the Frictional Coefficient}
\label{mu}
\begin{figure}[ht]
\begin{center}
\includegraphics[totalheight=0.17\textheight]{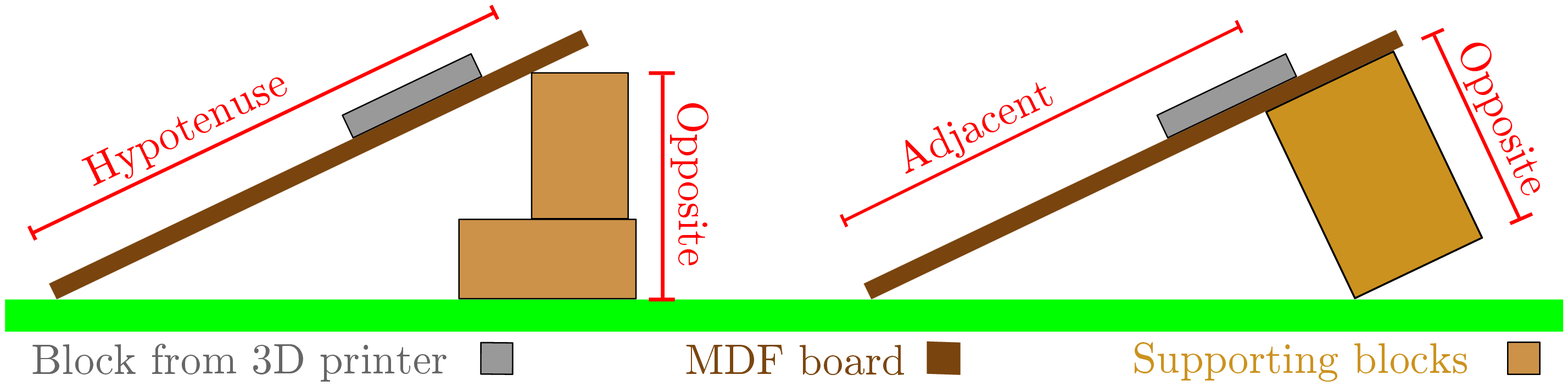}
\caption{Frictional Coefficient measures by observing the behaviour of blocks on inclined surface}
\label{FricDiag}
\end{center}
\end{figure}

The coefficient of dynamic friction between the printed block material and the MDF board was determined by inclining the board at an angle and noting the behaviour of the block.  The lengths of two sides of the right angled triangle were measured, as per figure \ref{FricDiag}, in order to determine the angle of the slope.

The dynamic frictional coefficient is the tangent of the slope angle where a moving block has a roughly 50:50 chance of slowing to a stop.  The static frictional coefficient is the tangent of the slope angle where a block placed on the slope has a roughly 50:50 chance of remaining stationary after the experimenter lets go of it.  These values were found to be $\mu_{static} = 0.50 \pm 0.03$ and $\mu_{dynamic} = 0.29 \pm 0.01 $ as can be seen in Table \ref{table}.

\begin{table}[ht]
\caption{* Dynamic Friction ($\mu$) = $0.29 \pm 0.01$ \ \ ** Static Friction = $0.50 \pm 0.03$ \\}
\centering
\scriptsize
\begin{tabular}{|c c c | c c | c c c | c }
\hline\hline
Adjacent & Opposite & Hypotenuse & Slope Angle& $\tan(\text{angle})$ & & Moving (or Tapped) & Stationary & \\ 
(cm) & (cm) & (cm) & ($^{\circ}$) & & & Block Behaviour & Block Behaviour& \ \\ [0.9ex] 
\hline
&$16.3\pm0.1$&$60.5\pm0.2$&$15.6\pm0.1$&$0.280\pm0.002$& & Mostly stops & Sticks & \\
&$16.3\pm0.1$&$58.6\pm0.2$&$16.2\pm0.1$&$0.290\pm0.002$& & 50:50 (between stopping & Sticks & * \\
&&&&&& or accelerating) & \\
&$16.3\pm0.1$&$56.3\pm0.2$&$16.8\pm0.1$&$0.302\pm0.002$& & Mostly doesn't stop & Sticks \\$71.8\pm0.1 \ $\ &$34.8\pm0.1$&&$25.9\pm0.1$&$0.485\pm0.002$&&Doesn't stop & Sticks \\
$69.3\pm0.1 \ $\ &$34.8\pm0.1$&&$26.7\pm0.1$&$0.502\pm0.002$&&Doesn't stop & Sticks 50:50& **\\
$65.2\pm0.1 \ $\ &$34.8\pm0.1$&&$28.1\pm0.1$&$0.534\pm0.002$&&Doesn't stop & Always slips \\
[1ex]
\hline
\end{tabular}
\label{table}
\end{table}

\section{Invariance of Outcome When Changing Length, $m$}
\label{LengthApen}

\begin{figure}[ht]
\begin{center}
\includegraphics[totalheight=0.3\textheight]{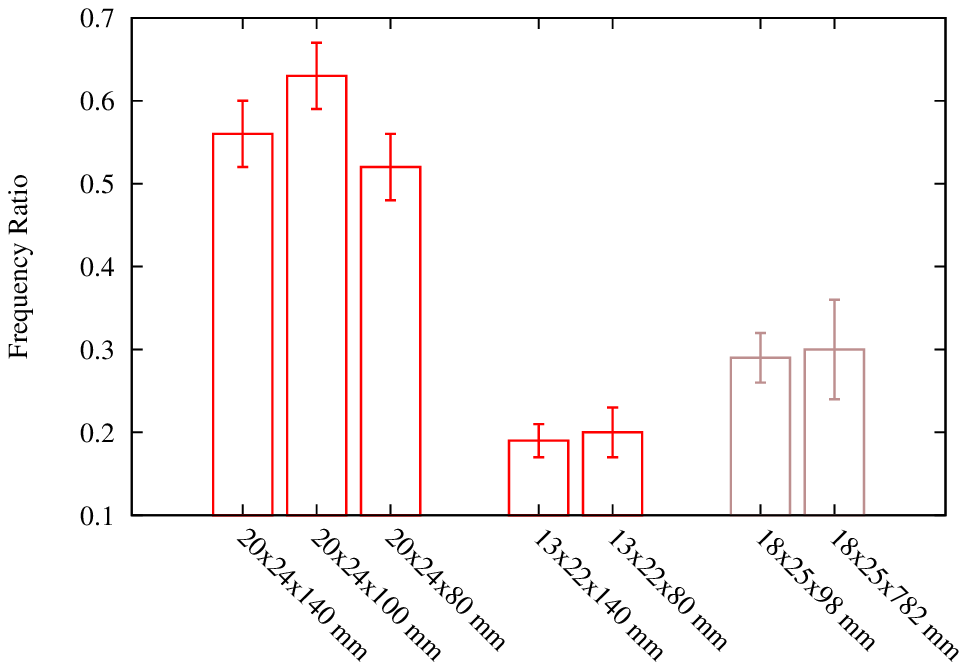}
\caption{Plot of the ratio of the number of Breadth Vertical outcomes to Width Vertical outcomes for three sets of blocks.  Within each set the Width and Breadth of the block was identical but the long length, $m$, varied.  The error bar shown is a binomial error.}
\label{LenGraph}
\end{center}
\end{figure}

To test that experimental blocks were sufficiently long such that the outcome of a roll was unaffected by the particular length of side $m$ we performed three sets of tests where $m$ was varied but the other dimensions were fixed.  Results from these tests in figure \ref{LenGraph} confirm our assertion.  We can conclude that, if the length is much longer than the other two-dimensions, it does not materially affect the outcome probabilities.  Full data in Table \ref{sidewaystable}, Section \ref{Experiment}.

\break

\section{Further Computational Discussion}
\label{CompApen}

The simulation code was written using the Bullet physics engine.  Bullet uses the symplectic Euler integration scheme \cite{integration} to integrate the equations of motions for sets of rigid bodies and impulses to resolve collisions between objects.  The simulation used a fixed integration timestep.  The size of this timestep was chosen such that we were confident that any further reduction in timestep would not materially affect the results.  More specifically, we are confident that the outcome frequency ratio was within one percent of the frequency ratio that would be yielded by the same simulation, with timestep reduced by a further order of magnitude. 

\begin{figure}[ht]
\begin{center}
\includegraphics[totalheight=0.3\textheight]{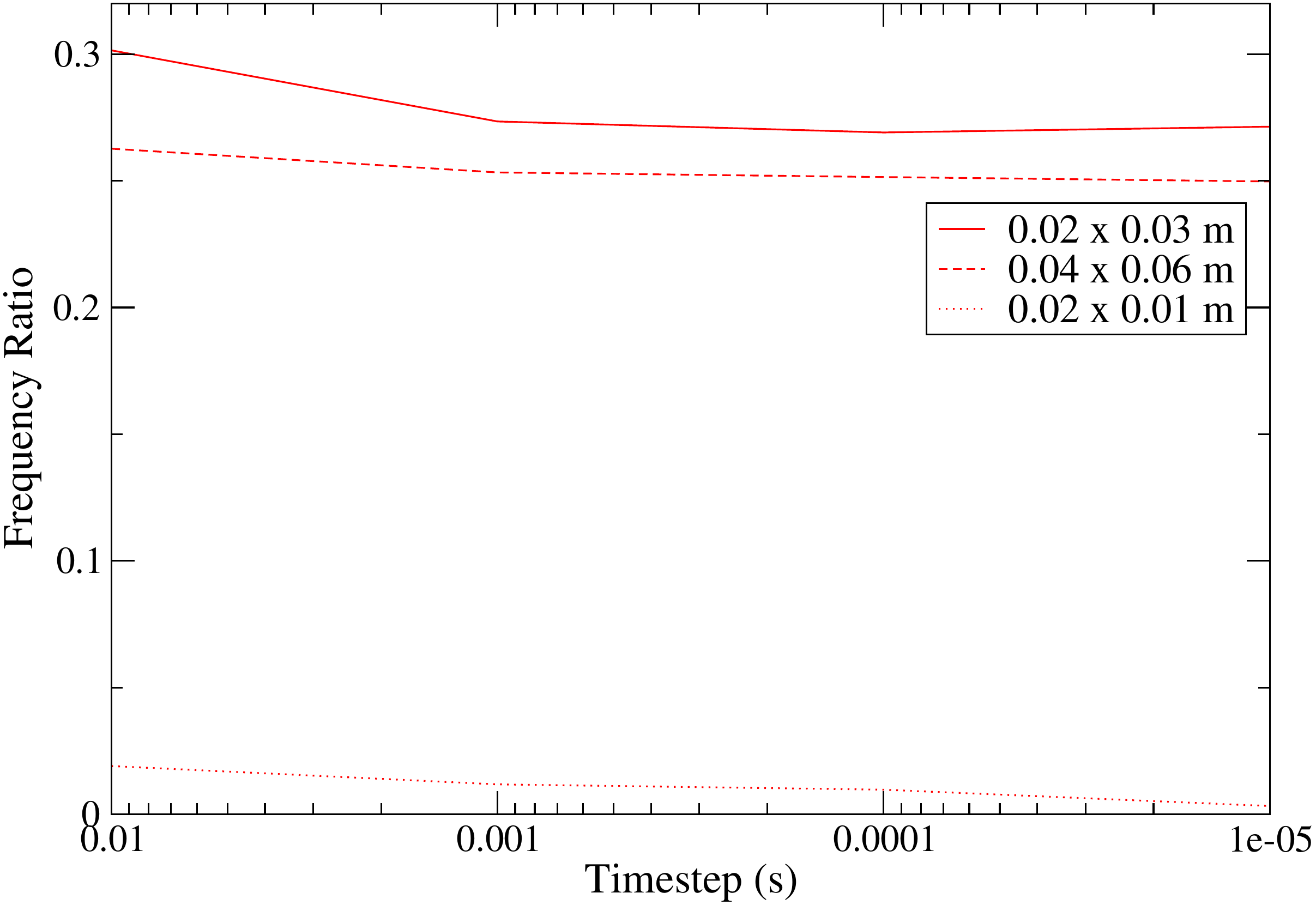}
\caption{Convergence testing for three block sizes showing how the rolling outcome is changed when decreasing the timestep.}
\label{convergence}
\end{center}
\end{figure}

Figure \ref{convergence} shows convergence test results for three blocks.  Based on these a timestep of 0.001 s was deemed to be sufficient.

\end{document}